\documentclass{mn2e}
\usepackage{graphicx}
\usepackage{amsbsy}
\usepackage{amsmath}
\usepackage{epsf}
\usepackage{paralist}
\usepackage{tablefootnote}

\newcommand{\apj}{ApJ}
\newcommand{\araa}{Annu. Rev. Astron. Astrophy}
\newcommand{\apjl}{ApJL}
\newcommand{\mnras}{MNRAS}
\newcommand{\apjs}{ApJS}
\newcommand{\pasp}{PASP}
\newcommand{\aap}{A\&A}

\newcommand{\icarus}{Icarus}

\newcommand{\beq}{\begin{equation}}
\newcommand{\eeq}{\end{equation}}

\DeclareMathAlphabet{\mathsfsl}{OT1}{cmss}{bx}{sl}
\SetMathAlphabet{\mathsfsl}{bold}{OT1}{cmss}{bx}{sl}

\usepackage{amssymb}
\usepackage{pifont}

\usepackage{hyperref}

\usepackage{epstopdf}
\usepackage{color}        

\begin{document}

\title[Simulating non-LTE line transfer]
{Python Radiative Transfer Emission code ($\textsc{PyRaTE}$): non-LTE spectral lines simulations}

\author[Tritsis et al.]
  {A.~Tritsis$^{1, 2}$, H.Yorke$^{3, 4}$, K.~Tassis$^{1,5}$ \\
    $^1$Department of Physics and ITCP \thanks{Institute for Theoretical and Computational Physics, formerly Institute for Plasma Physics}, University of Crete, PO Box 2208, 71003 Heraklion, Crete, Greece\\
    $^2$Research School of Astronomy and Astrophysics, Australian National University, Canberra, ACT 2611, Australia.\\
    $^3$Jet Propulsion Laboratory, California Institute of Technology, Pasadena, CA 91109, USA\\
    $^4$SOFIA Science Center, NASA Ames Research Center, 94035 Moffett Field, USA\\
    $^5$IESL and Institute of Astrophysics, Foundation for Research and Technology-Hellas, PO Box 1527, 71110 Heraklion, Crete, Greece}
\maketitle 

\begin{abstract}

We describe $\textsc{PyRaTE}$, a new, non-local thermodynamic equilibrium (non-LTE) line radiative transfer code developed specifically for post-processing astrochemical simulations. Population densities are estimated using the escape probability method. When computing the escape probability, the optical depth is calculated towards all directions with density, molecular abundance, temperature and velocity variations all taken into account. A very easy-to-use interface, capable of importing data from simulations outputs performed with all major astrophysical codes, is also developed. The code is written in $\textsc{Python}$ using an ``embarrassingly parallel'' strategy and can handle all geometries and projection angles. We benchmark the code by comparing our results with those from $\textsc{RADEX}$ (van der Tak et al. 2007) and against analytical solutions and present case studies using hydrochemical simulations. The code is available on $\textsc{GitHub}$.\footnotemark{}

\end{abstract}


\footnotetext{\url{https://github.com/ArisTr/PyRaTE}}
\section{Introduction}\label{intro}

Recent advancements in computer technologies, have allowed for simulations of multiphysics astrophysical problems to flourish. Numerical codes and methods are becoming increasingly sophisticated and accurate (Krumholz et al. 2007; Mocz et al. 2014a; Mocz et al.2014b; Schaal et al. 2015; Hopkins 2015; Hopkins \& Raives 2016). Astrochemical simulations in which the chemical and dynamical evolution of astrophysical objects are coupled, follow in the same direction (Tassis et al. 2012a; Tassis et al. 2012b; Clark et al. 2013; Motoyama et al. 2015; Walch et al. 2015; Tritsis et al. 2016; Seifried \& Walch 2016). 

Observational astrophysics is also advancing rapidly. New, very high angular and spectral resolution data obtained with telescopes such as ALMA can reveal features of astrophysical objects in unprecedented detail. Furthermore, such observations offer a unique opportunity to distinguish between different theoretical models. To do so, a connection between theoretical results and observations is essential. Since observationally our full range of knowledge about astrophysical objects is obtained through the study of electromagnetic radiation, a direct comparison requires post-processing simulation outputs with radiative transfer (RT) codes. 

Post-processing astrochemical simulations with line radiative transfer codes is not a trivial task. Local thermodynamic equilibrium (LTE) is not attained in many astrophysical problems and the population densities of the species under consideration have to be computed in non-local thermodynamic equilibrium (non-LTE). The majority of the existing radiative transfer codes (Keto 1990; Keto et al. 2004; Hogerheijde \& van der Tak 2000; van der Tak et al. 2007; Brinch \& Hogerheijde 2010; Dullemond 2012) do compute the population densities in non-LTE. However, the validity of certain approximations and assumptions, as for example is homogeneity and the calculation of optical depth from average quantities, may not always be appropriate. Furthermore, restrictions in the number of dimensions existing RT codes can handle and the fact that the interfaces are not always particularly practical can cause confusion and limit the number of choices a user has.

In this paper we describe the $\textsc{PyRaTE}$ (Python Radiative Transfer Emission) code. $\textsc{PyRaTE}$ is a user-friendly, ``embarrassingly parallel'', modular, non-LTE radiative transfer code, fully written in $\textsc{Python}$\footnote{\url{https://www.python.org/}}. It makes use of the $\textsc{yt}$ analysis toolkit (Turk et al. 2011) which can handle simulation outputs performed with all major codes\footnote{For a full list of the codes that $\textsc{yt}$ and thus $\textsc{PyRaTE}$ can handle see \url{http://yt-project.org/docs/dev/examining/loading_data.html}} as well as generic adaptive mesh refinement (AMR) and array data. As a result, exporting data from astrochemical simulations requires no effort. Moreover, $\textsc{PyRaTE}$ can handle all geometries and all projection angles. In its current version, the code is focused on star formation and molecular cloud related problems. However, it can be easily modified so that it can be used for any astrophysical problem.

In section \S~\ref{code} we describe the formalism of radiative transfer used and the numerical methods, we compare our results with those from $\textsc{RADEX}$ (van der Tak et al. 2007) and with analytical solutions, we outline the parallelism strategy followed and the design of the code. In \S~\ref{tests} we demonstrate the code's capabilities on hydrochemical numerical simulations. We use these simulations to demonstrate the code's capabilities. A summary of this work and future goals are given in \S~\ref{discuss}.

\section{code design}\label{code}
\subsection{Basic formalism}

Integration of the non-relativistic, time-independent equation of radiation transfer between two grid points $\textit{i}$ and $\textit{i}+1$ yields:
\begin{equation}\label{rt_eq}
I_{i+1}=\frac{(e^{-\tau_{i+1}^C}-p)I_i+pS_i^L+qS_{i+1}^L+S^k}{1+q}
\end{equation}  
(Yorke 1986) where the contributions of line and dust continuum emission, denoted with the superscripts $\textit{L}$ and $\textit{C}$ respectively, are considered separately. In Equation~\ref{rt_eq} $I$ is the radiative intensity, $S^L$ is the source function for line emission and $\tau^C$ is the optical depth for continuum emission. The quantities $q$, $p$ and $S^k$ are given by:
\begin{equation}\label{q}
q=\frac{\tau_{i+1}^L}{1+e^{-\tau_{i+1}^L}}
\end{equation}  
\begin{equation}\label{p}
p=q(e^{-\tau_{i+1}^L-\tau_{i+1}^C})
\end{equation}  
\begin{equation}\label{Sk}
S^k=e^{-\tau_{i+1}^C}\int_i^{i+1}\kappa^CS^Ce^{\int_i^s\kappa^C(s')ds'}ds
\end{equation}
where $S^C$ is the source function for dust continuum emission and $\tau_{i+1}^L$ is the optical depth of the line:
\begin{equation}\label{p}
\tau_{i+1}^L=\int_i^{i+1}\kappa^L(s)ds
\end{equation}  
In Equations~\ref{Sk} and~\ref{p} $s$ is the length along the line-of-sight (LOS), $\kappa^C$ is the extinction coefficient for continuum emission which can be computed from Mie theory (see Section ~\ref{dust}) and $\kappa^L$  is the line extinction coefficient:
\begin{equation}\label{kl} 
\kappa^L=n_mB_{mn}\frac{h\nu_0}{4\pi}\big[1-\frac{n_ng_m}{n_mg_n}\big]\varphi(\nu)
\end{equation}  
(Mihalas 1978). In the latter equation, $g$ is the statistical weight, $B$ is the Einstein B coefficient, $\nu_0$ is the rest frequency of the line, $h$ is the Planck constant, $n$ is the population density of the upper and lower level (denoted with the subscripts $n$ and $m$) and $\varphi(\nu)$ is the normalized profile function such that $\int \varphi(\nu) d\nu=1$. As a result, $I$ can be computed by calculating the quantities that appear in Equations~\ref{q} through~\ref{kl}, inserting them in Equation~\ref{rt_eq} and performing the ray-tracing.

\subsection{Dust model}\label{dust}

We use a three component dust grain model consisting of amorphous carbon grains, silicate grains and dirty ice (i.e. water, ammonia and carbon particles) coated silicate grains. Our grain model is that of Preibisch et al. (1993) which is appropriate for dense molecular cloud conditions. However, the amount of dust has been reduced to be compatible with the calculations performed by Pollack et al. (1985). The number of grains per gram gas are $logN_d$ = 14.170, 1. and 12.001 $\rm{gr^{-1}}$ for the amorphous carbon, silicate and dirty ice coated silicate grains respectively. The sublimation temperature of each component is $\sim$ 2000, 1500 and 125 K respectively. The total amount of amorphous carbon is kept constant by enhancing the number density of amorphous carbon grains in temperatures above 125 K. For every dirty ice coated silicate grain destroyed, a silicate grain is also added. All components are assumed to be spherical with sizes $\sim$ 10 $\rm{nm}$ for the carbon, $\sim$ 50 $\rm{nm}$ for the silicate and $\sim$ 60 $\rm{nm}$ for the dirty ice coated silicate grains. 
\begin{figure}
\includegraphics[width=1.0\columnwidth, clip]{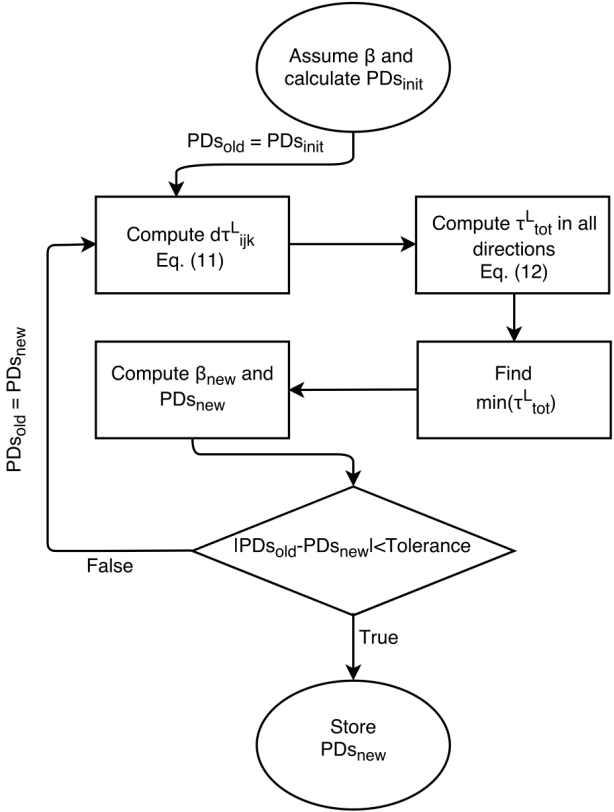}
\caption{Flow chart of the algorithm used to compute the population densities (PDs) at each grid point. The same procedure is followed for all grid points.
\label{algorithm}}
\end{figure}

\begin{figure}
\includegraphics[width=1.0\columnwidth, clip]{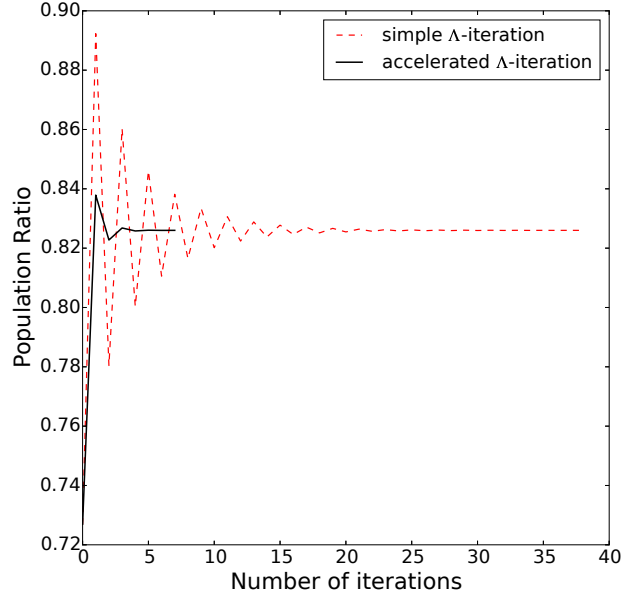}
\caption{The black solid line shows the performance of the accelerated $\Lambda$-iteration algorithm currently in use for calculating the population densities. The red dashed line show results using a simple $\Lambda$-iteration. The current algorithm converges in just a few iterations. Calculations were performed considering the optically thick $\rm{^{12}CO}$ molecule (J = 1 - 0 transition) and the physical parameters shown in Figure~\ref{sph_physical_params}.
\label{performance}}
\end{figure}

The data for the efficiency factor for extinction $Q^{ext}$ and the albedo $\alpha$ ($\alpha = Q^{sca}/Q^{ext}$, where $Q^{sca}$ is the efficiency factor for scattering) are stored in the code in tabulated form. Based on the desired frequency, linear interpolation is applied through the data for each of the dust components. The extinction coefficient for dust emission is then computed as:
\begin{equation}
\kappa^C=\sum\limits_{d=1}^3 n_dQ_d^{ext}\pi r_d^2
\end{equation}
where $n_d$ is the number density of each grain component which is in turn determined from the temperature and density of each grid point. The source function for continuum emission is computed as the sum of the thermal emission, weighted over the albedo, and the contribution from scattering. For the interpolation we use a standard $\textsc{scipy}$ routine\footnote{\url{http://docs.scipy.org/doc/scipy/reference/interpolate.html}}. Higher order interpolation through the data can also be used by changing a keyword in the interpolating function. For further details regarding our dust model and figures showing the absorption efficiencies as a function of frequency for each dust component we refer the reader to Preibisch et al. (1993) and Kessel et al. (1998). Finally, by adjusting this dust grain model, $\textsc{PyRaTE}$ can be modified to suit a variety of astrophysical problems.

\subsection{Computing the population ratio (non-LTE)}\label{algdescr}

We consider a multilevel system. The rate of change of the population density in level $m$ will be the sum of the excitation (absorption and collisional excitation) and de-excitation (induced emission, spontaneous emission and collisional de-excitation) processes:

\begin{equation}\label{SE}
\begin{split}
\frac{dn_m}{dt}=\sum_{n>m}^N[n_n(A_{nm}+B_{nm}u^{\nu}_{nm}+n_{H_2}C_{nm})\\-n_m(B_{mn}u^{\nu}_{mn}+n_{H_2}C_{mn})\big]+\\
\sum_{n<m}^N\big[-n_m(A_{mn}+B_{mn}u^{\nu}_{mn}+n_{H_2}C_{mn})\\+n_k(B_{nm}u^{\nu}_{nm}+n_{H_2}C_{nm}) \big]
\end{split}
\end{equation}
In Equation~\ref{SE}, $u^{\nu}_{nm}$ is the radiative energy density between levels $n$ and $m$, $C_{nm}$ and $C_{mn}$ (hereafter $n > m$) are respectively the de-excitation and excitation coefficients due to collisions, $\rm{n_{H_2}}$ is the number density of the $\rm{H_2}$ molecule which is the dominant collisional partner, and $A_{nm}$ and $B_{nm}$ are the Einstein A and B coefficients for spontaneous and induced emission. De-excitation coefficients are computed from the detailed balance relation:
\begin{equation}\label{de_exc_exc}
C_{mn}=C_{nm}\frac{g_n}{g_m}e^{-\Delta E/k_BT}
\end{equation}
where $k_B$ is the Boltzmann constant, $T$ is the temperature and $\Delta E$ is the energy difference between levels. In order to compute the terms containing the Einstein B coefficients on the right hand side of Equation~\ref{SE} a priori knowledge of the radiation field is required. As a result, the entire problem has to be solved iteratively, a computationally very demanding task. An alternative solution to the problem is to approximate $u_\nu$ as $S_\nu^L (1-\beta)$ where $\beta$ is the probability of a photon escaping the cloud. Following the same reasoning, the radiation field due to external photons penetrating the cloud will be, $S_\nu\beta$.

If we now further assume statistical equilibrium $\frac{dn_u}{dt}=0$ and take into account the conservation equation:
\begin{equation*}
\sum_m^N n_m = 1
\end{equation*}
we have a system of non-linear equations which $\textsc{PyRaTE}$ solves using an optimization routine from $\textsc{scipy}$ and which supports multiple methods for solution of non-linear systems\footnote{\url{https://docs.scipy.org/doc/scipy/reference/generated/scipy.optimize.root.html}}.

In order to determine the population densities we also have to compute the escape probabilities between radiatively coupled levels at every grid point. To do so, we use an accelerated $\Lambda$-iteration described in the following steps: 
\begin{enumerate}
\item{First, we assume some initial values for the escape probabilities (equal to 0.5 for all levels) and calculate the initial population densities for grid point ($i^\prime j^\prime k^\prime$) from Equation~\ref{SE}} and the conservation equation.\\ 
\item{Based on that initial guess for the population densities, the infinitesimal optical depth for all lines is computed at all grid points as:
\begin{equation}\label{dtau}
d\tau_{ijk}^L=\frac{c^3}{8\pi \nu_{0}^3}\frac{A_{nm}n_{ijk}^{molec}}{\Delta v_{ijk}^{th}}(\frac{g_n}{g_n}n_m-n_n)ds
\end{equation}
(van der Tak et al. 2007). In Equation~\ref{dtau}, $c$ is the speed of light, $n_{ijk}^{molec}$ is the number density of the molecule at grid point ($i$, $j$, $k$), $\Delta v_{th}$ is the thermal width of the line and $n>m$.}\\
\item{Next, the total optical depth for all lines is computed as the sum of the infinitesimal optical depth of the grid points for which their absolute velocity difference with grid point ($i^\prime j^\prime k^\prime$) is smaller than their thermal width.}\\
\item{The algorithm then searches the direction towards which the optical depths of the lines are minimum and the photons are more likely to escape the cloud (Poelman \& Spaans 2006; Perego et al. 2014; Scarlata \& Panagia 2015). For example, the minimum optical depth for a 3D cartesian grid is computed from six sums as:
\begin{eqnarray}\label{tau_poprat}
\tau_{i^\prime j^\prime k^\prime}^L= \min\Big(&& \sum\limits_{i=i^\prime+1}^Ad\tau_{ijk}^L[\mid v_{i^\prime j^\prime k^\prime}-v_{ij^\prime k^\prime}\mid<\Delta v_{ij^\prime k^\prime}^{th}], \\ \nonumber
&& \sum\limits_{i=i^\prime-1}^0d\tau_{ijk}^L[\mid v_{i^\prime j^\prime k^\prime}-v_{ij^\prime k^\prime}\mid<\Delta v_{ij^\prime k^\prime}^{th}],\\ \nonumber 
&& \sum\limits_{j=j^\prime+1}^Bd\tau_{ijk}^L[\mid v_{i^\prime j^\prime k^\prime}-v_{i^\prime jk^\prime}\mid<\Delta v_{i^\prime jk^\prime}^{th}],\\ \nonumber 
&& \sum\limits_{j=j^\prime-1}^0d\tau_{ijk}^L[\mid v_{i^\prime j^\prime k^\prime}-v_{i^\prime jk^\prime}\mid<\Delta v_{i^\prime jk^\prime}^{th}],\\ \nonumber 
&& \sum\limits_{k=k^\prime+1}^Cd\tau_{ijk}^L[\mid v_{i^\prime j^\prime k^\prime}-v_{i^\prime j^\prime k}\mid<\Delta v_{i^\prime j^\prime k}^{th}],\\ \nonumber 
&& \sum\limits_{k=k^\prime-1}^0d\tau_{ijk}^L[\mid v_{i^\prime j^\prime k^\prime}-v_{i^\prime j^\prime k}\mid<\Delta v_{i^\prime j^\prime k}^{th}]\nonumber \Big)
\end{eqnarray}
where $A$, $B$ and $C$ are the sizes of the grid in the $i$, $j$ and $k$ directions, respectively.}\\
\item{With that optical depths, new escape probabilities and population densities are computed.}\\
\item{If the relative change of the previously and newly computed population densities is smaller than some tolerance, then the values of the population densities are stored in computer memory. If this condition is not satisfied steps (ii) through (vi) are repeated until the population densities converge to some values.} 
\end{enumerate}

A flow chart of the algorithm that summarizes these steps is shown in Figure~\ref{algorithm}. With this procedure, density, temperature, velocity and molecular abundance variations are all taken into account. After the population densities have been computed, the line source function is calculated as:
\begin{equation}
S^L=\frac{2h\nu^3}{c^2}\frac{1}{\frac{n_mg_n}{n_ng_m}-1}
\end{equation}

The described procedure is a simple $\Lambda$-iteration. To increase the computational speed we implement an accelerated scheme. In each successive iteration, the new population densities are a weighted mean of the two previously computed values instead of just the previously computed value. Furthermore, since escape probabilities are not expected to vary much between adjacent grid points, results for one grid point are used as initial guesses for the neighbouring cell. In Figure~\ref{performance} we show an example where the population ratio for a grid point is calculated using simple (red dashed line) and accelerated (black solid line) $\Lambda$-iterations. In the latter case the algorithm converges within a few iterations. Both tests were performed using the 1D isothermal spherical collapse model (the physical parameters of which are shown in Figure~\ref{sph_physical_params}) with 304 grid points, a tolerance of $10^{-5}$ and the optically thick $\rm{^{12}CO}$ molecule.

\begin{figure}
\includegraphics[width=1.0\columnwidth, clip]{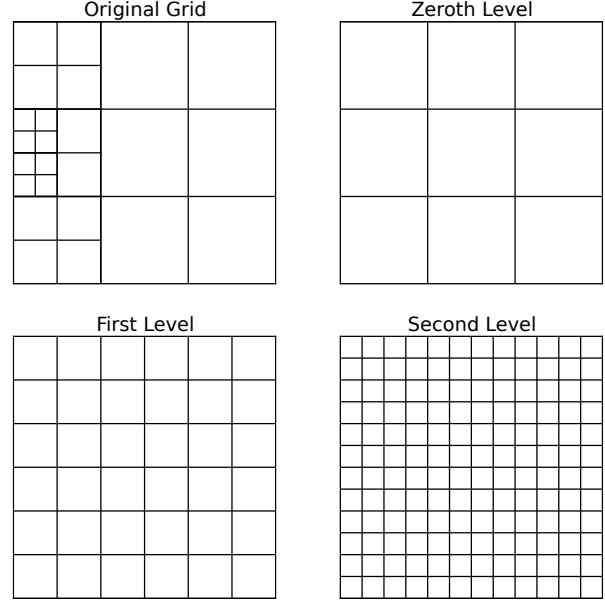}
\caption{Treatment of AMR simulation outputs. Depending on the desired resolution, from the original AMR grid, a new uniform grid is created. When the resolution of the maximum level of refinement is selected, new grid points are created with values equal to the one of the grid point of the parent level.  
\label{amr_treatment}}
\end{figure}

\begin{figure}
\includegraphics[width=1.02\columnwidth, clip]{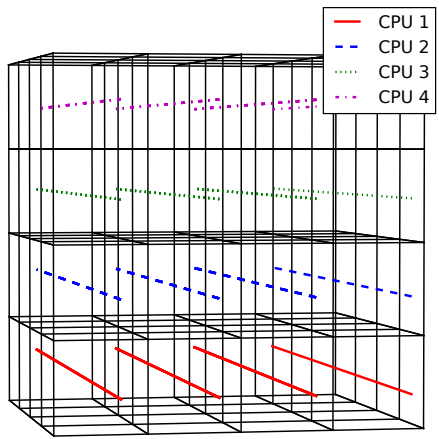}
\caption{Strategy followed to parallelize the algorithms in our code. During ray tracing each processor is responsible only for a set of rays (coloured lines) passing through the computational region. 
\label{parallelization}}
\end{figure}

\begin{figure}
\includegraphics[width=1.0\columnwidth, clip]{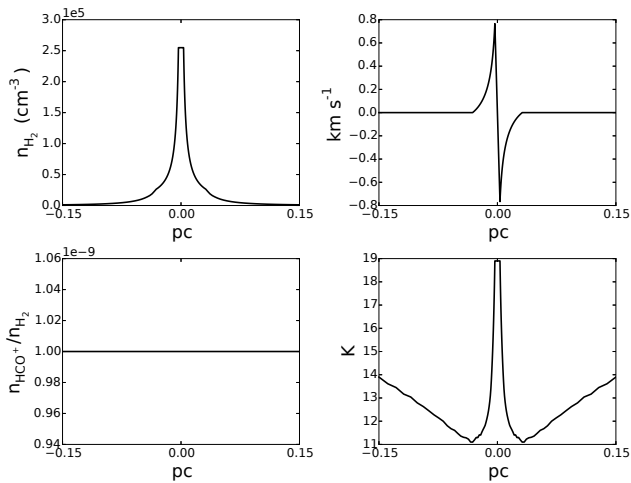}
\caption{Parameters of the spherical model adopted from $\textsc{RADEX}$'s test problem web page (van Zadelhoff et al. 2002) to investigate the accuracy of the algorithm developed here for computing the population ratio. In the upper left panel we plot the density, in the upper right the velocity, in lower left the molecular abundance and in the lower right the temperature.
\label{model_for_populations}}
\end{figure} 

\subsection{Loading data and constructing the grid}

Data from astrochemical simulations and from files containing molecular data\footnote{Molecular data are taken from the Leiden Atomic and Molecular Database (Sch{\"o}ier et al. 2005) \url{http://home.strw.leidenuniv.nl/~moldata/}} are initially loaded into the code and stored in computer memory in tabulated form. These include the mass density, the abundance of the molecule under consideration, the Einstein coefficients, collisional excitation/de-excitation coefficients, the dimensions of the grid and velocities in all directions. Magnetic fields for Zeeman splitting measurements will also be included in a future version. 

When loading data we assume spherical symmetry for 1D simulation outputs, plane and axial symmetries for 2D simulation outputs in cylindrical geometry and no symmetries for 3D simulations. These symmetries are accounted for in our calculations. However, the module responsible for exporting simulation data can be easily modified such that the desired symmetries and velocity conventions are followed. Where necessary, volume rendering is performed simultaneously with ray tracing. Throughout the code, we use cgs units. Velocities with a minus sign represent motions towards the observer and vice versa. All other conventions followed in the code (e.g. conversion of frequencies to velocities) are consistent with the ones followed in radio astronomy. 

Simulation outputs performed both in adaptive mesh refinement (AMR) and in uniform grids are supported. However, AMR grids are re-sampled into uniform grids based on the resolution of a refinement level. In the upper left panel of Figure~\ref{amr_treatment} we show an example of an AMR grid with two levels of refinement. In the upper right, lower left and lower right panels we show the resulting grids constructed from the original grid, based on the zeroth, first and second level of refinement respectively. The values of additional grid points created in the lower panels of Figure~\ref{amr_treatment} have the same values as the ones of their parent level. This treatment has the advantage of simplifying the algorithms and making them easy to read and modify. However, adding more grid points can substantially increase the computational time, especially for AMR simulations with many levels of refinement. In a future version we plan to upgrade the treatment of AMR grids.

\subsection{Parallelization}\label{parallel}

In order to reduce computational time we have parallelized the majority of the code's modules using an ``embarrassingly parallel'' strategy. ``Embarrassingly parallel'' or naturally parallel problems are problems that can be decomposed in identical subtasks all of which can be solved independently without the need for communication between processors. In Figure~\ref{parallelization} we show an example of a 3D grid in which ray tracing is performed on four CPUs. Each CPU is responsible only for a set of rays (coloured lines in Figure~\ref{parallelization}). Similarly, in the modules computing the population ratios, the grid is split over the number of available processors and is then reconstructed. Given this strategy the code is linearly scalable. $\textsc{PyRaTE}$ uses the package $\textsc{mpi4py}$\footnote{\url{https://pypi.python.org/pypi/mpi4py}} which provides functions and bindings for the Message Passing Interface (MPI) and which can be easily installed alongside $\textsc{yt}$. 

\subsection{Auxiliary modules}

A number of modules are available for pre-processing the simulation data and post-processing the results after ray-tracing is performed. The user has the option to add microturbulent broadening:
\begin{equation}
\Delta v_{tot}=\sqrt{\Delta v_{turb}^2+\Delta v_{th}^2}
\end{equation}
where $\Delta v_{th}$ is the thermal linewidth and $\Delta v_{turb}$ may be a constant number for all grid points or random numbers drawn from a flat or a Gaussian distribution. Additional modules also exists for debugging purposes and for creating simple models so there is no need for a priori astrochemical simulations. 

To facilitate comparison with observations the user has a number of options. Gaussian noise can be added to spectral lines, based on a user-defined signal-to-noise ratio (SNR) and emission maps can be convolved with Gaussian filters based on some desired distance and telescope's beam size. Spectral lines can be plotted in antenna or brightness temperature, mJy or in cgs units. Furthermore, lines can be smoothed to match the spectral resolution of observations. The resulting output format of $\textsc{PyRaTE}$ is equivalent to a position-position-velocity (PPV) cube from observations which can also be saved in FITS format for further analysis. Options for creating integrated emission maps are also available.

\subsection{Benchmarking}

\begin{figure}
\includegraphics[width=1.0\columnwidth, clip]{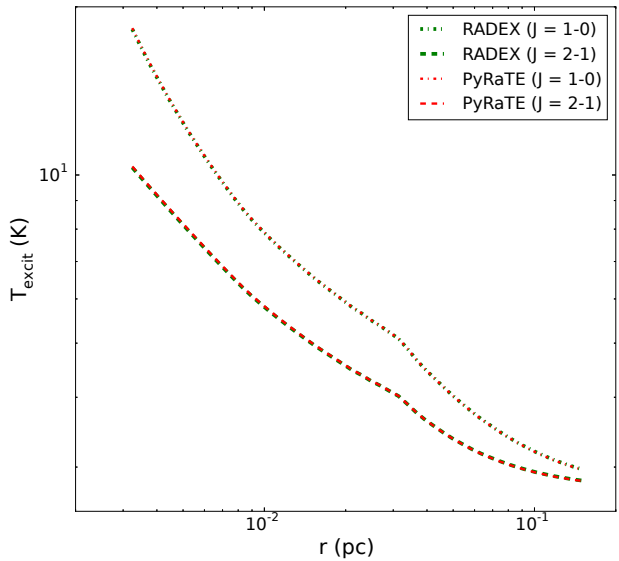}
\caption{Comparison of the excitation temperatures obtained with $\textsc{PyRaTE}$ (dashed-dotted and dashed red lines) and $\text{RADEX}$ (dashed-dotted and dashed green lines). When we follow the same approximations with $\text{RADEX}$ results obtained with the two codes are identical. The parameters of the model used to compare the population densities between the two codes are shown in Figure~\ref{model_for_populations}.
\label{poprat_comparison}}
\end{figure}

In order to test the validity of the algorithm developed we compare our results with those of $\textsc{RADEX}$ (van der Tak et al. 2007). To do so, we adopt the set-up of a test-problem of $\textsc{RATRAN}$ (Hogerheijde \& van der Tak 2000; van Zadelhoff et al. 2002), the parameters of which are shown in Figure~\ref{model_for_populations}. In this model, the abundance of $\rm{HCO^{+}}$ is low ($\sim 10^{-9}$), so that the emission is optically thin, and the radial profiles of the physical parameters were created based on the model by Shu (1977). Results for the excitation temperature for J = 1 - 0 and J = 2 - 1 obtained with both codes, are shown in Figure~\ref{poprat_comparison} and are identical. For producing Figure~\ref{poprat_comparison}, instead of computing the optical depth as described in section~\ref{algdescr} we have made the same approximations as in $\textsc{RADEX}$, namely that the optical depth can be computed from average quantities as:

\begin{equation}\label{dtau}
d\tau^L=\frac{c^3}{8\pi \nu_{0}^3}\frac{A_{nm}N^{molec}}{1.064\Delta V}(\frac{g_n}{g_n}n_m-n_n)
\end{equation}
where $\Delta V$ is the full width at half-maximum (FWHM) of the line and $N^{molec}$ is the total molecular column density both of which are given as a model parameters. This formalism leads to an overestimation of the optical depth, low escape probabilities and thus more molecules in excited states.

\begin{figure}
\includegraphics[width=1.0\columnwidth, clip]{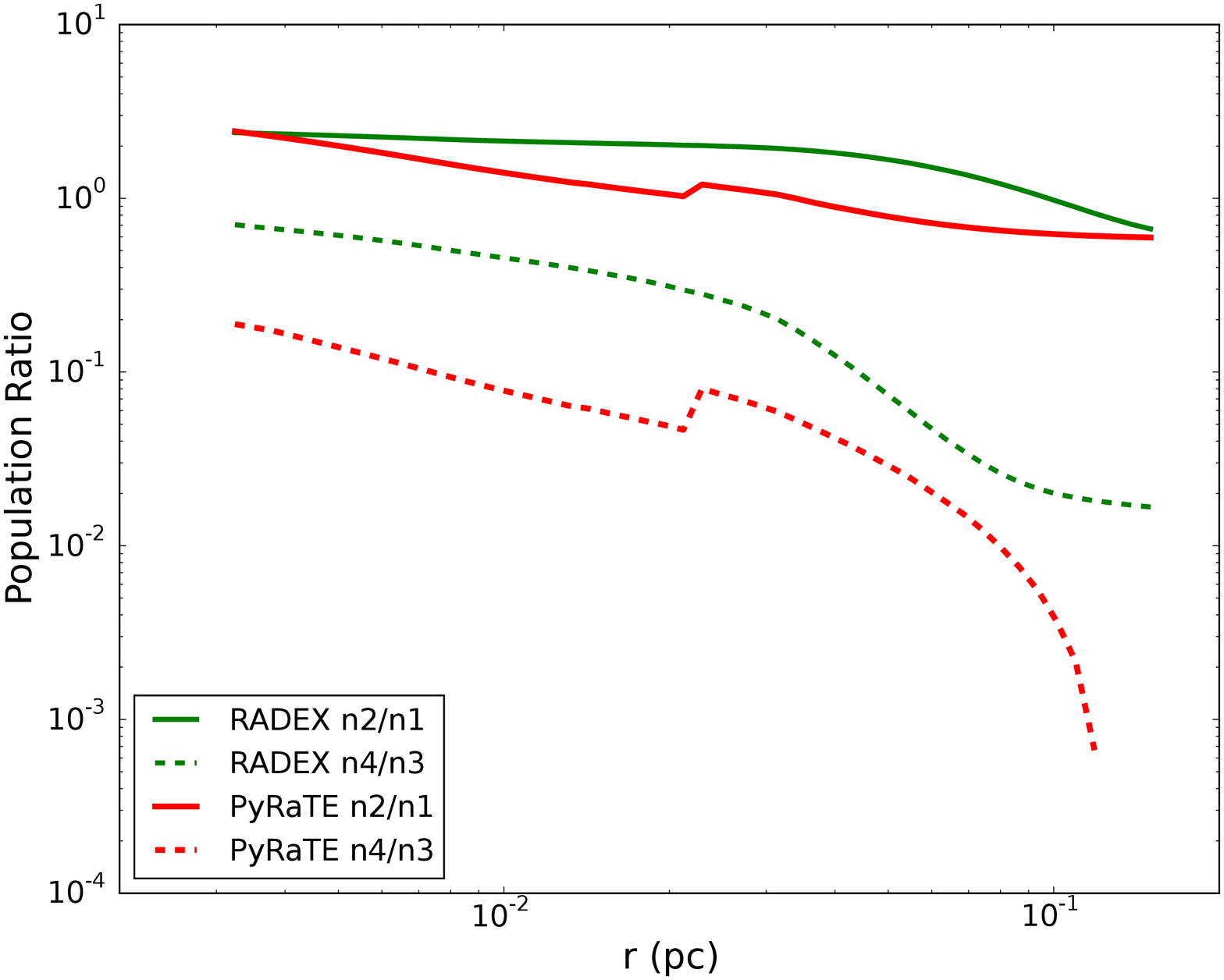}
\caption{Comparison of population densities obtained with $\textsc{PyRaTE}$ (solid and dashed red lines) and $\text{RADEX}$ (solid and dash-dotted green lines) for the same physical model as in Figure~\ref{poprat_comparison} but with $10^2$ higher $\rm{HCO^+}$ abundance. Using the algorithm described in~\ref{algdescr} the optical depth computed with $\textsc{PyRaTE}$ is smaller and thus less molecules are in excited states.
\label{poprat_comparison2}}
\end{figure}

In order to demonstrate this effect we plot in Figure~\ref{poprat_comparison2} the population ratios obtained with $\textsc{RADEX}$ and $\textsc{PyRaTE}$ for the same physical model. However, $\rm{HCO^+}$ abundance is increased by $10^2$. For calculating the optical depth we have used the algorithm described in section~\ref{algdescr}. Because of the fact that the optical depth computed with $\textsc{PyRaTE}$ is lower than that computed from average quantities there is not enough photon trapping and thus less molecules are in excited states. The kinks in the population ratios (solid and dashed red lines) are due to abrupt changes in the escape probabilities. In turn, this abrupt change is caused by the velocity profile (see Figure~\ref{model_for_populations}). Beyond a radius of $\sim$ 0.025 pc the absolute velocity difference between grid points is always less than the thermal width and as a result the optical depth rises. Thus, the kinks shown in Figure~\ref{poprat_comparison2} are because of the adopted test problem and $\textit{not}$ because of numerical resolution effects. The same behaviour is not seen when the abundance is $10^{-9}$ and the emission is optically thin everywhere.

\begin{figure}
\includegraphics[width=1.0\columnwidth, clip]{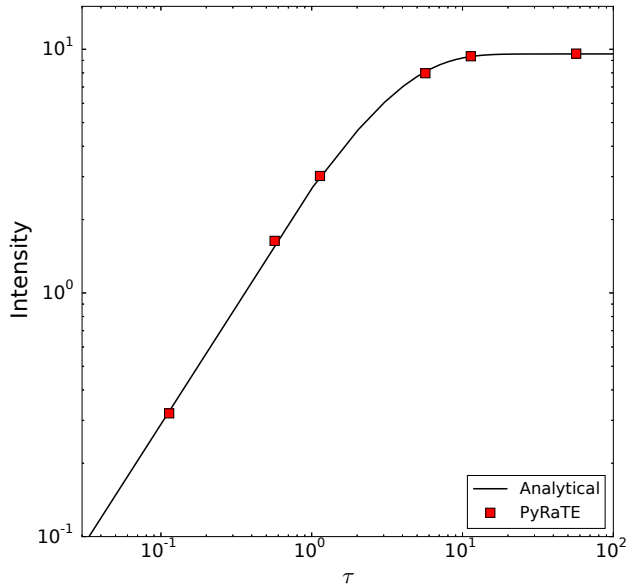}
\caption{The intensity at the centre of the line $\rm{CO} ~(1 - 0)$ as a function of the ``mean" optical depth, as this is defined in Kylafis (1983). With the solid black line we plot the results from the analytical solution and with the red squares the results from the code. The intensity is plotted in units of $h\nu^3/c^2$.
\label{analytical}}
\end{figure}

To further test the developed algorithms we compare results from $\textsc{PyRaTE}$ against the analytical solutions derived in Kylafis (1983) for an one dimensional problem applicable to $\rm{CO}$. In this problem, the velocity field is of the form $\vec{v}=(v_0/L)z\hat{k}$ and the temperature, $\rm{CO}$ abundance and density are uniform everywhere. In Figure~\ref{analytical} we show the intensity at the centre of the line (J = 1 - 0) as this is calculated from the analytical expressions (solid black line) and numerically (red squares) as a function of the "mean" optical depth defined in Kylafis (1983). Results from the code are in perfect agreement with the analytical method.

\section{Test cases}\label{tests}

The numerical simulations used to demonstrate $\textsc{PyRaTE}$'s capabilities were performed with the astrophysical code FLASH4.0.1 (Fryxell et al. 2000, Dubey et al. 2008) and were described in detail in Tritsis et al. (2016). Here we give a brief overview of these simulations. In Tritsis et al. (2016) we employed hydrodynamic simulations of prestellar cores in spherical and cylindrical symmetry. The radius of our spherically symmetric models was 0.55 pc. The axial and radial dimensions of one of our cylindrically symmetric models were 0.72 and 0.48 pc respectively. The initial number density was $\rm{10^3~cm^{-3}}$. In all models, all velocity components were initially set to zero and the cores were allowed to collapse freely under their self-gravity. The dynamical models were isothermal and coupled with non-equilibrium chemical network consisting of 214 gas-phase and 82 dust grain species with each species treated as a different fluid. Our chemical network consisted of $\sim$ 14000 reactions with reaction rates adopted from the fifth release of the $\textsc{UMIST}$ database (McElroy et al. 2013).

\begin{figure}
\includegraphics[width=1.0\columnwidth, clip]{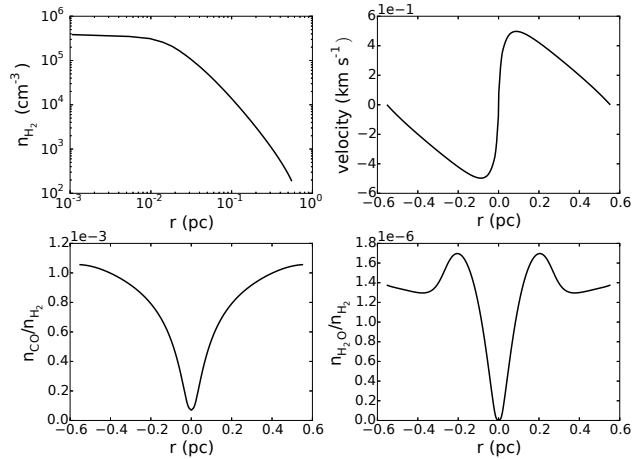}
\caption{Physical and chemical parameters used to produced the synthetic lines shown in Figure~\ref{co_example_sph} and Figure~\ref{absorption}. The parameters are taken from a hydrodynamical simulation of a collapsing isothermal sphere. The upper left and right panels show the radial profiles of number density and velocity respectively. In the lower left panel we show the $\rm{CO}$ abundance profile and in the lower right that of $\rm{H_2O}$. 
\label{sph_physical_params}}
\end{figure}

\begin{figure}
\includegraphics[width=1.0\columnwidth, clip]{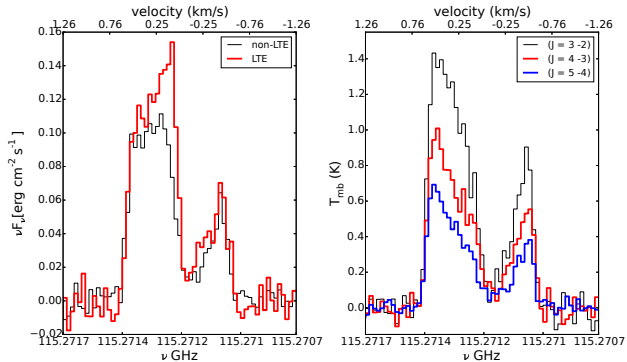}
\caption{Left panel: $^{12}$CO (J = 1 - 0) emission line from a ray passing through the centre of a simulated spherical core under the LTE approximation (bold red line) and with non-LTE (black line). Right panel: excited-state lines for the same ray. Gaussian noise with a signal-to-noise ratio of $\sim$ 5 was added to the lines. The code correctly reproduces the expected infall asymmetry in the line profiles.
\label{co_example_sph}}
\end{figure}

In Figure~\ref{sph_physical_params} we show the density (upper left), velocity (upper right), $\rm{CO}$ abundance (lower right) and $\rm{H_2O}$ abundance distribution for our spherical model when the central density is 4$\cdot~10^5~\rm{cm^{-3}}$. The temperature is constant and equal to 7 K everywhere. This evolutionary stage corresponds to a time of $\sim$ 1 Myrs. In the left panel Figure~\ref{co_example_sph} we show a $^{12}$CO (J = 1 - 0) emission line from a LOS threading the middle of our spherical core. With the red line we plot our results assuming LTE and the black line non-LTE calculations. The spectral resolution used to produce the synthetic lines was $\sim$ 0.05 km/s. Random, gaussian noise was added to the line with a signal-to-noise ratio (SNR) of $\sim$ 5. $\textsc{PyRaTE}$ correctly reproduces a line profile characteristic of infall motions (Evans 1999). The emission line is double-peaked with the blue component being stronger due to absorption along the LOS. The strength of the line changes depending on whether LTE is assumed. Since in non-LTE we expect that the upper level will be less populated than in LTE, the source function of the line, and thus the intensity, will decrease. This feature is clearly seen in Figure~\ref{co_example_sph}, especially for the blue shifted component of the line since for the red shifted component the decrease in the source function is counteracted by the decrease in the absorption coefficient. Furthermore, from Figure~\ref{co_example_sph} it can be seen that the line profile changes depending on whether LTE is assumed. That is due to the velocity profile. The velocity differences between the centre of the core and the outer parts are larger than the thermal width of the line. Consequently, the total optical depth decreases, the escape probability increases and $\rm{CO}$ is only subthermally excited. However, if we were to ``observe" the two lines, taking resolution restrictions into account, both profiles could probably be fitted by two Gaussian functions. In the right panel of Figure~\ref{co_example_sph} we show excited-state lines for the same ray as in the left panel.

\begin{figure}
\includegraphics[width=1.0\columnwidth, clip]{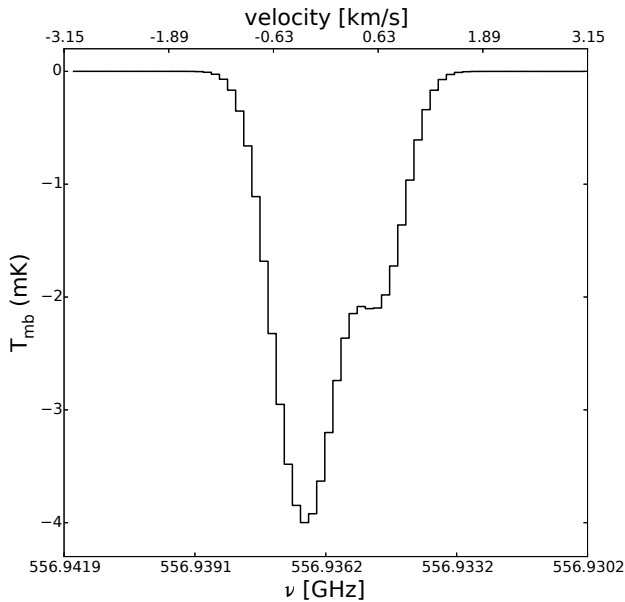}
\caption{$\rm{H_2O}$ ($\rm{1_{10}}$ - $\rm{1_{01}}$) absorption line from a ray passing through the centre of our simulated spherical core. The continuum level has been subtracted. The line was produced using a spectral resolution of 0.1 km/s and was then smoothed to a spectral resolution of 0.6 km/s. 
\label{absorption}}
\end{figure}

In Figure~\ref{absorption} we show a synthetic absorption line of $\rm{H_2O}$ ($\rm{1_{10}}$ - $\rm{1_{01}}$) from a LOS threading the middle of our spherical core. The spectral resolution was 0.1 km/s but the line was then smoothed to a spectral resolution of 0.6 km/s. The continuum level was $\sim$ 11 mK (i.e. the same as in Caselli et al. 2010 for L1544) and was subtracted. The CO (J = 1 - 0) emission line combined with the absorption line for $\rm{H_2O}$ for the same physical parameters is in agreement with expectations from both observations and previous line radiative transfer simulations (Caselli et al. 2010; Caselli et al. 2012).

\begin{figure*}
\includegraphics[width=2.0\columnwidth, clip]{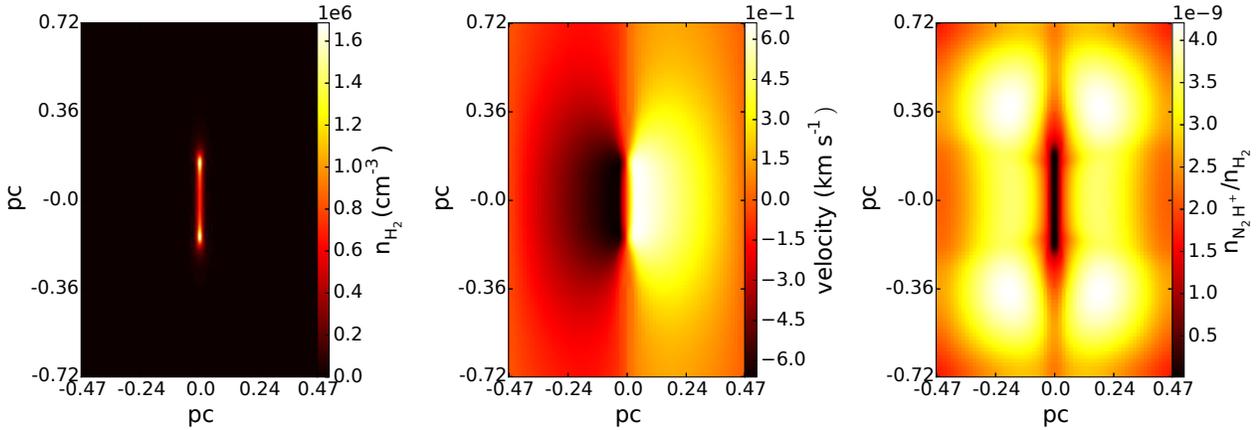}
\caption{Physical and chemical parameters used to produced the synthetic PPV cube shown in Figure~\ref{cyl_iso_collapse}. The parameters are taken from a hydrodynamical simulation of a collapsing isothermal cylinder. The left, middle and right panels show the number density, radial velocity and $\rm{N_2H^{+}}$ abundance distribution respectively.
\label{cyl_physical_params}}
\end{figure*}

\begin{figure*}
\includegraphics[width=2.0\columnwidth, clip]{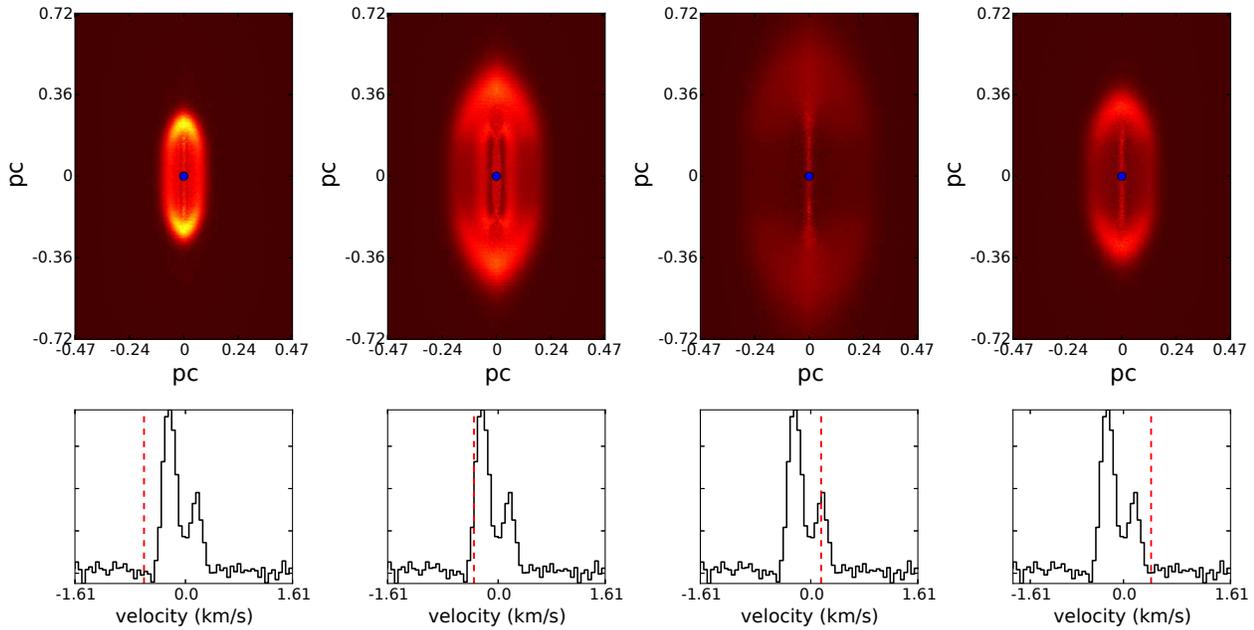}
\caption{Four velocity slices (upper row) from a synthetic PPV cube of a collapsing isothermal cylinder seen in $\rm{N_2H^{+}}$ (J = 1 - 0). The lower row shows the spectrum from a LOS passing through the centre of the map (blue points in upper row) while the red line traces the velocity of each slice map.
\label{cyl_iso_collapse}}
\end{figure*}

In order to demonstrate the capabilities of $\textsc{PyRaTE}$ to post-process simulations with more complex geometries we also produce a synthetic PPV cube from hydrodynamical simulations of a collapsing isothermal cylindrical/filamentary core. In Figure~\ref{cyl_physical_params} we show the physical parameters of the core. In the left panel we show the number density distribution, in the middle panel we plot the radial velocity and in the right panel we show the $\rm{N_2H^+}$ abundance distribution. The temperature is equal to 10 K. In Figure~\ref{cyl_iso_collapse} we show four velocity slices from our synthetic PPV cube of the core as observed in $\rm{N_2H^+}$ (J = 1 - 0) and with a spectral resolution of $\sim$ 0.05 km/s. In the upper panel we plot the emission map in each velocity slice and in the lower panel a spectral line (black line) threading the centre of the map (blue points in the upper panels). The red line in the lower panel traces the velocity shown in the upper panel. The core is observed such that its axis of symmetry is perpendicular to the LOS (edge-on).

\section{Summary}\label{discuss}

Radiative transfer is essential when comparing results from astrochemical simulations with radio observations. To this end, we have developed an easy-to-use non-LTE line radiative transfer code. The code can be used to post-process results from astrochemical simulations performed with all major astrophysical codes and can handle all geometries and projection angles. The population densities are computed using the escape probability method with variations in density, molecular abundance, temperature and velocities taken into account. 

Compared to existing radiative transfer codes the populations densities in $\textsc{PyRaTE}$ are $\textit{not}$ computed from average quantities and velocity variations are taken into account. Its simple interfaces both for importing and exporting data and the fact that it is written in $\textsc{Python}$ make the code easy-to-use, flexible and capable of producing publication quality figures effortlessly.

In a future version we plan to include a full treatment of the Stokes parameters in our code. Thus, simulations of the polarization of spectral line emission will also become possible. The algorithms for ray-tracing and for computing the population densities will continue to be developed both in terms of performance and accuracy. We think that $\textsc{PyRaTE}$ can be proven an important tool in the effort of connecting theory with observations. The code is publicly available for the community at \url{https://github.com/ArisTr/PyRaTE}.

\section*{Acknowledgements}

We thank N. Kylafis for useful comments and discussions. We also thank the anonymous referee for useful comments that helped improved this paper. The software used in this work was in part developed by the DOE NNSA-ASC OASCR Flash Center at the University of Chicago. For post processing our results we partly used yt analysis toolkit (Turk et al. 2011). K.T. and A.T. acknowledge support by FP7 through Marie Curie Career Integration Grant PCIG- GA-2011-293531 ``SFOnset". A.T. and K.T. would like to acknowledge partial support from the EU FP7 Grant PIRSES-GA-2012-31578 ``EuroCal". Usage of the Metropolis HPC Facility at the CCQCN of the University of Crete, supported  by  the European Union Seventh Framework Programme (FP7-REGPOT-2012-2013-1) under grant agreement no. 316165, is also acknowledged.


\begin{thebibliography}{99}
\bibitem[Brinch \& Hogerheijde(2010)]{2010A&A...523A..25B} Brinch, C., \& Hogerheijde, M.~R.\ 2010, \aap, 523, A25
\bibitem[Caselli et al.(2010)]{2010A&A...521L..29C} Caselli, P., Keto, E., Pagani, L., et al.\ 2010, \aap, 521, L29  
\bibitem[Caselli et al.(2012)]{2012ApJ...759L..37C} Caselli, P., Keto, E., Bergin, E.~A., et al.\ 2012, \apjl, 759, L37 
\bibitem[Clark et al.(2013)]{2013ApJ...768L..34C} Clark, P.~C., Glover, S.~C.~O., Ragan, S.~E., Shetty, R., \& Klessen, R.~S.\ 2013, \apjl, 768, L34
\bibitem[Dubey et al.(2008)]{2008ASPC..385..145D} Dubey, A., Fisher, R., Graziani, C., et al.\ 2008, Numerical Modeling of Space Plasma Flows, 385, 145 
\bibitem[Dullemond(2012)]{2012ascl.soft02015D} Dullemond, C.~P.\ 2012, Astrophysics Source Code Library, ascl:1202.015 
\bibitem[Evans(1999)]{1999ARA&A..37..311E} Evans, N.~J., II 1999, \araa, 37, 311 
\bibitem[Fryxell et al.(2000)]{2000ApJS..131..273F} Fryxell, B., Olson, K., Ricker, P., et al.\ 2000, \apjs, 131, 273
\bibitem[Goldsmith \& Li(2005)]{2005ApJ...622..938G} Goldsmith, P.~F., \& Li, D.\ 2005, \apj, 622, 938 
\bibitem[Hogerheijde \& van der Tak(2000)]{2000A&A...362..697H} Hogerheijde, M.~R., \& van der Tak, F.~F.~S.\ 2000, \aap, 362, 697 
\bibitem[Hopkins(2015)]{2015MNRAS.450...53H} Hopkins, P.~F.\ 2015, \mnras, 450, 53 
\bibitem[Hopkins \& Raives(2016)]{2016MNRAS.455...51H} Hopkins, P.~F., \& Raives, M.~J.\ 2016, \mnras, 455, 51
\bibitem[Kessel et al.(1998)]{1998A&A...337..832K} Kessel, O., Yorke, H.~W., \& Richling, S.\ 1998, \aap, 337, 832 
\bibitem[Keto(1990)]{1990ApJ...355..190K} Keto, E.~R.\ 1990, \apj, 355, 190 
\bibitem[Keto et al.(2004)]{2004ApJ...613..355K} Keto, E., Rybicki, G.~B., Bergin, E.~A., \& Plume, R.\ 2004, \apj, 613, 355 
\bibitem[Krumholz et al.(2007)]{2007ApJ...667..626K} Krumholz, M.~R., Klein, R.~I., McKee, C.~F., \& Bolstad, J.\ 2007, \apj, 667, 626 
\bibitem[Kylafis(1983)]{1983ApJ...267..137K} Kylafis, N.~D.\ 1983, \apj, 267, 137 
\bibitem[McElroy et al.(2013)]{2013A&A...550A..36M} McElroy, D., Walsh, C., Markwick, A.~J., et al.\ 2013, \aap, 550, A36 
\bibitem[Mihalas(1978)]{1978stat.book.....M} Mihalas, D.\ 1978, San Francisco, W.~H.~Freeman and Co., 1978.~650 p.,   
\bibitem[Mocz et al.(2014a)]{2014MNRAS.437..397M} Mocz, P., Vogelsberger, M., Sijacki, D., Pakmor, R., \& Hernquist, L.\ 2014a, \mnras, 437, 397
\bibitem[Mocz et al.(2014b)]{2014MNRAS.442...43M} Mocz, P., Vogelsberger, M., \& Hernquist, L.\ 2014b, \mnras, 442, 43 
\bibitem[Motoyama et al.(2015)]{2015ApJ...808...46M} Motoyama, K., Morata, O., Shang, H., Krasnopolsky, R., \& Hasegawa, T.\ 2015, \apj, 808, 46
\bibitem[Perego et al.(2014)]{2014A&A...568A..11P} Perego, A., Gafton, E., Cabez{\'o}n, R., Rosswog, S., \& Liebend{\"o}rfer, M.\ 2014, \aap, 568, A11 
\bibitem[Poelman \& Spaans(2006)]{2006A&A...453..615P} Poelman, D.~R., \& Spaans, M.\ 2006, \aap, 453, 615  
\bibitem[Pollack et al.(1985)]{1985Icar...64..471P} Pollack, J.~B., McKay, C.~P., \& Christofferson, B.~M.\ 1985, \icarus, 64, 471
\bibitem[Preibisch et al.(1993)]{1993A&A...279..577P} Preibisch, T., Ossenkopf, V., Yorke, H.~W., \& Henning, T.\ 1993, \aap, 279, 577 
\bibitem[Scarlata \& Panagia(2015)]{2015ApJ...801...43S} Scarlata, C., \& Panagia, N.\ 2015, \apj, 801, 43 
\bibitem[Schaal et al.(2015)]{2015MNRAS.453.4278S} Schaal, K., Bauer, A., Chandrashekar, P., et al.\ 2015, \mnras, 453, 4278 
\bibitem[Sch{\"o}ier et al.(2005)]{2005A&A...432..369S} Sch{\"o}ier, F.~L., van der Tak, F.~F.~S., van Dishoeck, E.~F., \& Black, J.~H.\ 2005, \aap, 432, 369 
\bibitem[Seifried \& Walch(2016)]{2016MNRAS.459L..11S} Seifried, D., \& Walch, S.\ 2016, \mnras, 459, L11
\bibitem[Shirley(2015)]{2015PASP..127..299S} Shirley, Y.~L.\ 2015, \pasp, 127, 299 
\bibitem[Shu(1977)]{1977ApJ...214..488S} Shu, F.~H.\ 1977, \apj, 214, 488 
\bibitem[Tassis et al.(2012a)]{2012ApJ...753...29T} Tassis, K., Willacy, K., Yorke, H.~W., \& Turner, N.~J.\ 2012a, \apj, 753, 29
\bibitem[Tassis et al.(2012b)]{2012ApJ...754....6T} Tassis, K., Willacy, K., Yorke, H.~W., \& Turner, N.~J.\ 2012b, \apj, 754, 6
\bibitem[Tritsis et al.(2016)]{2016MNRAS.458..789T} Tritsis, A., Tassis, K., \& Willacy, K.\ 2016, \mnras, 458, 789 
\bibitem[Turk et al.(2011)]{2011ApJS..192....9T} Turk, M.~J., Smith, B.~D., 
Oishi, J.~S., et al.\ 2011, \apjs, 192, 9 
\bibitem[van der Tak et al.(2007)]{2007A&A...468..627V} van der Tak, F.~F.~S., Black, J.~H., Sch{\"o}ier, F.~L., Jansen, D.~J., \& van Dishoeck, E.~F.\ 2007, \aap, 468, 627 
\bibitem[van Zadelhoff et al.(2002)]{2002A&A...395..373V} van Zadelhoff, G.-J., Dullemond, C.~P., van der Tak, F.~F.~S., et al.\ 2002, \aap, 395, 373 
\bibitem[Walch et al.(2015)]{2015MNRAS.454..238W} Walch, S., Girichidis, P., Naab, T., et al.\ 2015, \mnras, 454, 238
\bibitem[Yorke(1986)]{1986ASIC..188..141Y} Yorke, H.~W.\ 1986, NATO Advanced Science Institutes (ASI) Series C, 188, 141 


\end{thebibliography}
\end{document}